\documentclass[prb, twocolumn, superscriptaddress]{revtex4-2}
\usepackage{amsmath,amssymb,bm}
\usepackage{hyperref}
\usepackage{graphicx}
\usepackage{epstopdf}
\usepackage{latexsym}
\usepackage[usenames, dvipsnames]{color}
\usepackage[usenames, dvipsnames]{xcolor}
\usepackage{braket}
\usepackage{float}
\usepackage[normalem]{ulem}
\usepackage{comment}
\usepackage{mathtools}
\usepackage{array}
\usepackage{tabu}
\usepackage{multirow}
\usepackage[inline]{enumitem}
\usepackage{cleveref}
\usepackage{xcolor}
\usepackage[normalem]{ulem}
\usepackage{natbib}
\usepackage{amsmath}

\begin{document}
\title{The effect of disorder on phases across two-dimensional thermal melting}
\author{Prashanti Jami}
\affiliation{Department of Physics, Indian Institute of Science Education and Research (IISER) Kolkata, Mohanpur - 741246, West Bengal, India}
\author{Pinaki Chaudhuri}
\affiliation{The Institute of Mathematical Sciences, Taramani, Chennai 600113, India}
\author {Chandan Dasgupta}
\affiliation{Department of Physics, Indian Institute of Science, Bangalore 560012, India}
\affiliation{International Centre for Theoretical Sciences, Tata Institute of Fundamental Research, Bangalore 560089, India}
\author{Amit Ghosal}
\affiliation{Department of Physics, Indian Institute of Science Education and Research (IISER) Kolkata, Mohanpur - 741246, West Bengal, India}

\begin{abstract}
We study melting in a two-dimensional system of classical particles with Gaussian-core interactions in disordered environments. The pure system validates the conventional two-step melting with a hexatic phase intervening between the solid and the liquid. This picture is modified in the presence of pinning impurities. A random distribution of pinning centers forces a hexatic-like low temperature phase that
transits into a liquid at a single melting temperature $T^{\rm RP}_{\rm m}$. In contrast, pinning centers located at randomly chosen sites of a perfect crystal anchors a solid at low temperatures which undergoes a direct transition to the liquid at $T^{\rm CP}_{\rm m}$. Thus, the two-step melting is lost in either cases of disorder.
We discuss the characteristics of melting depending on the nature of the impurities.
\end{abstract}

\maketitle 

\begin{figure}[b]
\includegraphics[width=0.42\textwidth]{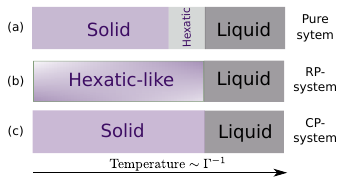}
\caption{A schematic representation of 2D-melting in: (a) KTHNY description in a pure system, (b) A system with randomly placed impurities, and (c) With a fraction of particles frozen at  randomly chosen zero-temperature positions of the pure system, suggesting that the clean-melting may be obscured by impurities (See text).}
\label{f0}
\end{figure}

\textit{Introduction \textemdash }
Enhanced fluctuations make two-dimensional melting a topic of immense research interest.
Unlike their three-dimensional counterparts undergoing ``Lindemann melting" \cite{lindemann1910calculation, Lozovik_1987}, 2D melting is mediated by the unbinding of topological defects.
The positional order (PO) and bond-orientational order (BOO) decouple in 2D, producing a ``hexatic phase" sandwiched between the solid and the liquid. Hexaticity, a rich concept, is realized in colloids~\cite{PhysRevLett.82.2721, cphc.200900755}, the vortex lattice in superconductors~\cite{Guillamon2009}, in active Brownian disks~\cite{PhysRevLett.121.098003}
and recently in van der Waals magnet~\cite{Meisenheimer2023}.
The celebrated KTHNY theory~\cite{J_Kosterlitz_1972, Kosterlitz:1973xp, PhysRevLett.41.121, PhysRevB.19.2457, PhysRevB.19.1855}, pictures 2D-melting as a two-step process involving successive unbinding of dislocations and disclinations-- presented schematically in Fig.~\ref{f0}(a). 
However, the relevance of the two-step 2D melting has also been  debated~\cite{PhysRevLett.114.035702, PhysRevB.28.178, C4SM00125G,Mazars_2015}.

Quenched disorder, inherent to real materials, can not only move around the phase boundaries but is also capable of modifying the mechanism of melting. For example, impurities can generate unbounded defects even at $T=0$, and thereby mask the unbinding of thermal defect-pairs. This could strike out solidity even at the lowest $T$, as suggested by Nelson~\cite{PhysRevB.27.2902} -- portrayed schematically in Fig.~\ref{f0}(b).   
In contrast, impurities which pin a given fraction of particles on sites of the underlying perfect lattice could stabilize the solid by anchoring it via these commensurate locations, and thereby consume the phase space of hexaticity (see Fig.~\ref{f0}(c)). Role of disorder in destabilizing the hexatic phase ~\cite{C4SM02876G} and in enhancing long-range correlation has also been pointed out ~\cite{Guillamon2014}. Thus, a careful analysis of 2D-melting in disordered media can potentially uncover new paradigms.

Experiments on
colloids~\cite{PhysRevLett.111.098301}, vortex lattices~\cite{PhysRevB.93.144503, duhan2023structure} and multi component mixtures~\cite{PhysRevLett.130.258202} indicate a broadened stability of the hexatic phase in the presence of disorder consistent with recent calculations~\cite{ PhysRevE.103.062612, MD177, Shankaraiah_2020, H_2020, PhysRevE.92.032110,PhysRevLett.74.4867}. Study of 2D melting is also popular in confined geometry~\cite{ctpp.201200028,PhysRevE.96.042105,PhysRevE.98.042134} mimicking disordered background. 
Zeng {\it et al.}\cite{PhysRevLett.82.1935} 
have argued that a solid (``Bragg glass'') phase with power-law decay of translational correlations cannot occur in a 2D system with impurities. Pronounced hexatic correlations are expected to be present \cite{PhysRevB.40.11355, PhysRevB.43.7831}
if the disorder is not strong, though there are controversies \cite{PhysRevLett.66.2523, PhysRevLett.67.1809, PhysRevLett.67.1810}
about the existence of a hexatic glass phase with long- or quasi-long-range hexatic order in 2D.
In contrast, phase boundaries of the two-step melting are found to be insensitive to quenched defects on a spherical surface~\cite{singh2022observation}. 

In this letter, we investigate the phases across melting of a bulk 2D system of soft-core particles, modeled via Gaussian-core interactions~\cite{10.1063/1.432891},
which is known to validate the KTHNY-melting scenario in a pure system ~\cite{PhysRevLett.106.235701}. 
Addressing the role of quenched disorder in the phase behaviour this model, our key results are summarized as follows:
(i) Random-pinning (RP) destabilizes solidity causing a single transition from a low-$T$ hexatic-like phase to a high-$T$ liquid. Here, the low-$T$ phase undergoes a likely crossover from hexatic-glass to hexatic-liquid. (ii) On the other hand, the commensurate-pinning (CP) anchors solidity and engulfs hexaticity -- even the high-$T$ liquid phase supports inhomogeneous pockets of crystallinity. The defect locations correlate oppositely with pinning centers in the two models of disorder -- defects tend to bind with the pinning centers for RP-systems, whereas they stay away from the impurities in CP-systems. Thus, in either realization of the quenched disorder, the two-step melting is lost.

\textit{Model \& method: \textemdash }
We introduce disorder in two different ways: (a) Random pinning (RP), in which we freeze a given fraction ($n_{\rm imp}$) of particles, chosen randomly in space, within a high-$T$ liquid configuration. Here, these immobile particles act as disorder.
(b) Commensurate pinning (CP), where $n_{\rm imp}$ fraction of particles are frozen at randomly chosen positions of an ideal triangular lattice -- the ground state configuration of the pure system. Note that CP represents correlated disorder with a long-range positional correlation of a perfect lattice. In contrast, RP constitutes nearly uncorrelated disorder though weak short-range correlation of a high-$T$ liquid may exist.
We investigated a system of $N=4356$ particles, with $n_{\rm imp}= 3.5\%$.
These results were compared with those from a pure system with $N=4096$ particles.
For these systems, we sample configurations via molecular dynamics.~\cite{frenkel2001understanding}
 within the canonical ensemble, using LAMMPS~\cite{PLIMPTON19951}  
 We consider a simulation box having dimensions $L_{x}= \frac{2}{\sqrt{3}} L_{y}$, having periodic boundary conditions. $L_{x}$ is adjusted to keep the density, $\rho$ of particles fixed for all our studies ($\rho = 0.628$).
We carried out $2 \times 10^7$ MD steps with a time step $\delta t = 0.005$. 
 We use dimensionless parameters: $t' = t \sqrt{\epsilon/m \sigma^2}$ and $E' = E/\epsilon$, where $m$ is the mass of each particle. $T$ is expressed in terms of $\Gamma^{-1}$\cite{PhysRevB.20.326}, where $\Gamma = \epsilon \exp(-\sqrt{3}/2\rho)/K_{B}T$.
The physical observables are averaged over 
$8$~-~$10$ independent pinning configurations for a given  $n_{\rm imp}$.

\textit{Positional and bond orientational order: \textemdash }
A pure 2D solid is characterized by two kinds of ordering: (i) PO  measured by $\psi_{\rm T} = \frac{1}{N}\left\langle|\Psi_{\rm T} |\right\rangle$, where  $\Psi_{\rm T} =  \sum_{i=1}^{N} \exp(i{\bf G}. {\bf r}_{i})$. %
$\bf {G}$ is a first shell reciprocal-lattice vector of the underlying triangular crystal and ${\bf r}_i$ is the position of particle $i$, and (ii) BOO, quantified by 
$\psi_{\rm 6}= \frac{1}{N}\left\langle| \sum_{k=1}^{N}\Psi_{\rm 6}(r_{k}) |\right\rangle$ where
$\Psi_{\rm 6} = \frac{1}{N_{b}(k)}\sum_{l=1}^{N_{b}(k)}
\exp{(i6\theta_{kl})} $. 
The sum is over the $N_b(k)$ nearest neighbors of particle $k$ identified by a Voronoi construction~\cite{TIPPER1991597} and $\theta_{kl}$ is the angle that a line joining particle $k$ and particle $l$ makes with a reference axis.
KTHNY theory predicts two critical temperatures:  $\Gamma^{-1}_{\rm SH}$ and  $\Gamma^{-1}_{\rm HL}$, for the thermal depletion of quasi-long-range PO (solid to hexatic) and BOO (hexatic to liquid) respectively, leaving a hexatic phase with quasi-long-range BOO between the solid and isotropic liquid phases.

\begin{figure}[t]
\includegraphics[width=0.48\textwidth]{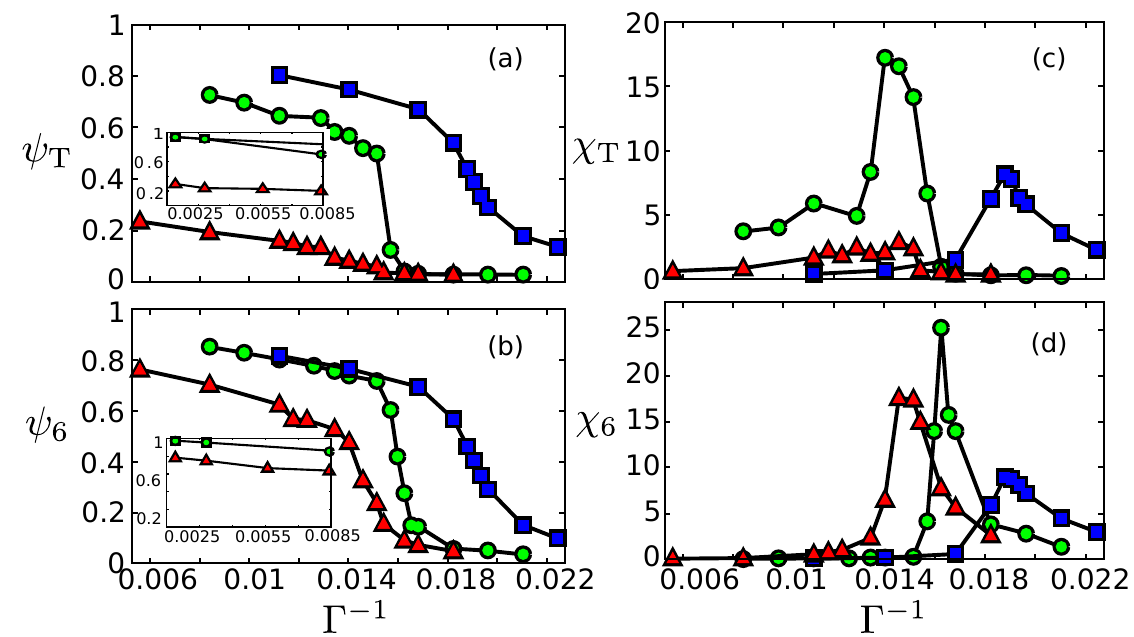}
\caption{
{\bf Positional and orientational ordering} tendencies are shown as a function of $\Gamma^{-1}$ for pure (green circles), RP (red triangles) and CP (blue square) systems. 
(a) Decay of PO ($\psi_{\rm T}$) with $\Gamma^{-1}$
(the inset shows $\psi_{\rm T}$ for $\Gamma^{-1} \rightarrow 0$).
(b) The softening of the BOO ($\psi_{\rm 6}$). Here $\psi_{\rm 6}(\Gamma^{-1}\rightarrow 0)$ is shown in inset. Panel (c) and (d) show the corresponding susceptibilities $\chi_{\rm T}$ and $\chi_{\rm 6}$. The location of the peaks in $\chi_{\rm T}$ and $\chi_{\rm 6}$ identify the transitions.
We find  $\Gamma^{-1}_{\rm RP} = 0.0145$ from $\chi_{\rm 6}$. Interestingly, the peak in $\chi_{\rm T}$ and $\chi_{\rm 6}$ for CP-systems appear at the same $\Gamma^{-1}_{\rm CP} = 0.0187$. }
\label{f1}
\end{figure}

In Fig.~\ref{f1}(a,b), we plot the thermal evolution of $\psi_{\rm T}$ and $\psi_{\rm 6}$ for pure, RP and CP systems. While the pure system follows KTHNY melting~\footnote{The three phases are clearly identified by the snapshots at three representatives $T$s in the SM, Fig.~S1(a-c)} with  $\Gamma^{-1}_{\rm SH} = 0.0140$ and $\Gamma^{-1}_{\rm HL} = 0.0162$, $\psi_{\rm T}$ in the CP-system is found to survive to larger $T$. The RP-system shows a much weaker $\psi_{\rm T}$ than the other two, even at the lowest $T$ and depletes very gradually with $T$ without any threshold behavior.
A threshold behavior near $\Gamma^{-1}_{\rm HL}$ is also seen in the pinned systems in Fig.~\ref{f1}(b), albeit the transitions are broader. Unlike $\psi_{\rm T}$, the $\psi_{\rm 6}$ is comparable at low-$T$ in pure, CP- and RP-systems. We also note that $\psi_{\rm T}$ and $\psi_{\rm 6}$ show a significant drop at the same critical temperature in a CP-system, implying a direct transition from solid to liquid, which we discuss further below.

In addition, the fluctuations of $\psi_{\rm T}$ and $\psi_{\rm 6}$ define  generalised susceptibilities $\chi_\alpha = \frac{1}{N}\left[\langle \vert\Psi_\alpha\vert^2\rangle-\langle \vert\Psi_\alpha\vert\rangle^2\right]$, (with $\alpha=T$ or $6$), and help to identify 
$\Gamma^{-1}_{\rm SH}$ and  $\Gamma^{-1}_{\rm HL}$, as shown in Fig.~\ref{f1}(c,d). Their behavior confirms that the pure system shows sharp transitions. Consistent with our finding in panel (a), $\chi_{\rm T}$ in the RP-system features only a broad and low hump, hinting that a low-$T$ phase in such a system represents a broad crossover between a hexatic glass~\cite{PhysRevB.40.11355} and a hexatic liquid ~\footnote{The low-$T$ phase support significant hexatic order $\psi_{\rm 6}$ for our model parameters, while $\psi_{\rm T}$ and the snapshots establish the amorphous and glassy nature. The $\Gamma^{-1}$-dependence of $\psi_{\rm 6}$ and $\psi_{\rm T}$ is indicative of a crossover from a hexatic-glass to a hexatic-liquid, before the RP-system transits to a liquid at $\Gamma^{-1}_{\rm RP}$. However, the resolution of our simulation is inadequate for drawing a firm conclusion.}. This is also consistent with the trajectory picture of RP system at $\Gamma^{-1} = 0.0028$ in the supplementary material (SM) ~\cite{f_note} Fig.~S1(d).
Congruous with our findings in panels (a,b), the locations of the peak of $\chi_{\rm T}$ and $\chi_{\rm 6}$ verify that PO and BOO in the CP-system vanish at a single $\Gamma^{-1}_{\rm CP}$. 
While our results from Fig.~\ref{f1} seem to support the schematic phase-diagram of Fig.~\ref{f0}, we emphasize that the `impure' phases at low- and high-$T$ defy conventional wisdom. These include the presence of unbound defects even at $T=0$ in RP-systems, and pockets of crystallinity deep into the liquid phase in CP-systems, as seen from Fig.~S1(g-i) in the SM ~\cite{f_note}, and discussed below.

\begin{figure}[t]
    \includegraphics[width=0.48\textwidth]{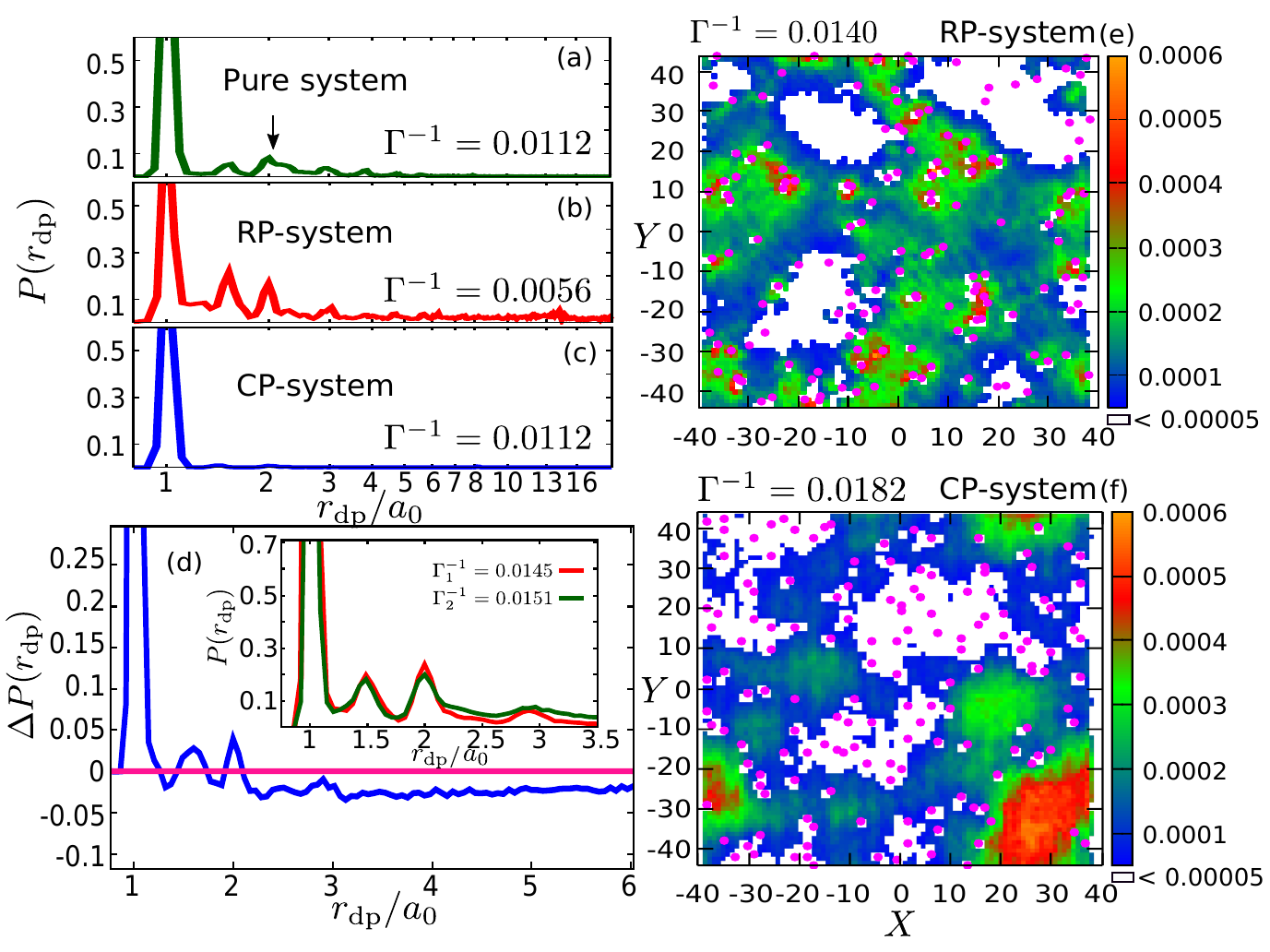}
    \caption{
    {\bf The distribution of distances between dislocation pairs,} $P(r_{\rm dp})$ in different phases. Panels (a)-(c) show results for pure, RP- and CP-systems, respectively, at low-$T$. $a_{0}$ is the average inter-particle distance. $P(r_{\rm dp})$ is sharply peaked at the typical distance between dislocation pairs, though there are differences in details.
    The distribution has a short tail for the pure system, a very long tail for RP-systems, and essentially no tail for CP-systems. 
   Panel (d) displays $\Delta P(r_{\rm dp})$ taken in pure system between $T's$ just above and below $\Gamma^{-1}_{\rm SH}$. The inset shows corresponding $P(r_{\rm dp})$. The largest zero-crossing distance of $\Delta P(r_{\rm dp})$ is taken as $r^c_{\rm dp}$, and is marked as an arrow in panel (a).
    Panels (e) and (f) show the spatial density of unbound defects just prior to melting into a liquid for a given realization of RP- and CP-system. The pinning centers are marked via magenta dots.}
    \label{f2}
\end{figure}

\begin{figure}[t]
    \includegraphics[width=0.48\textwidth]{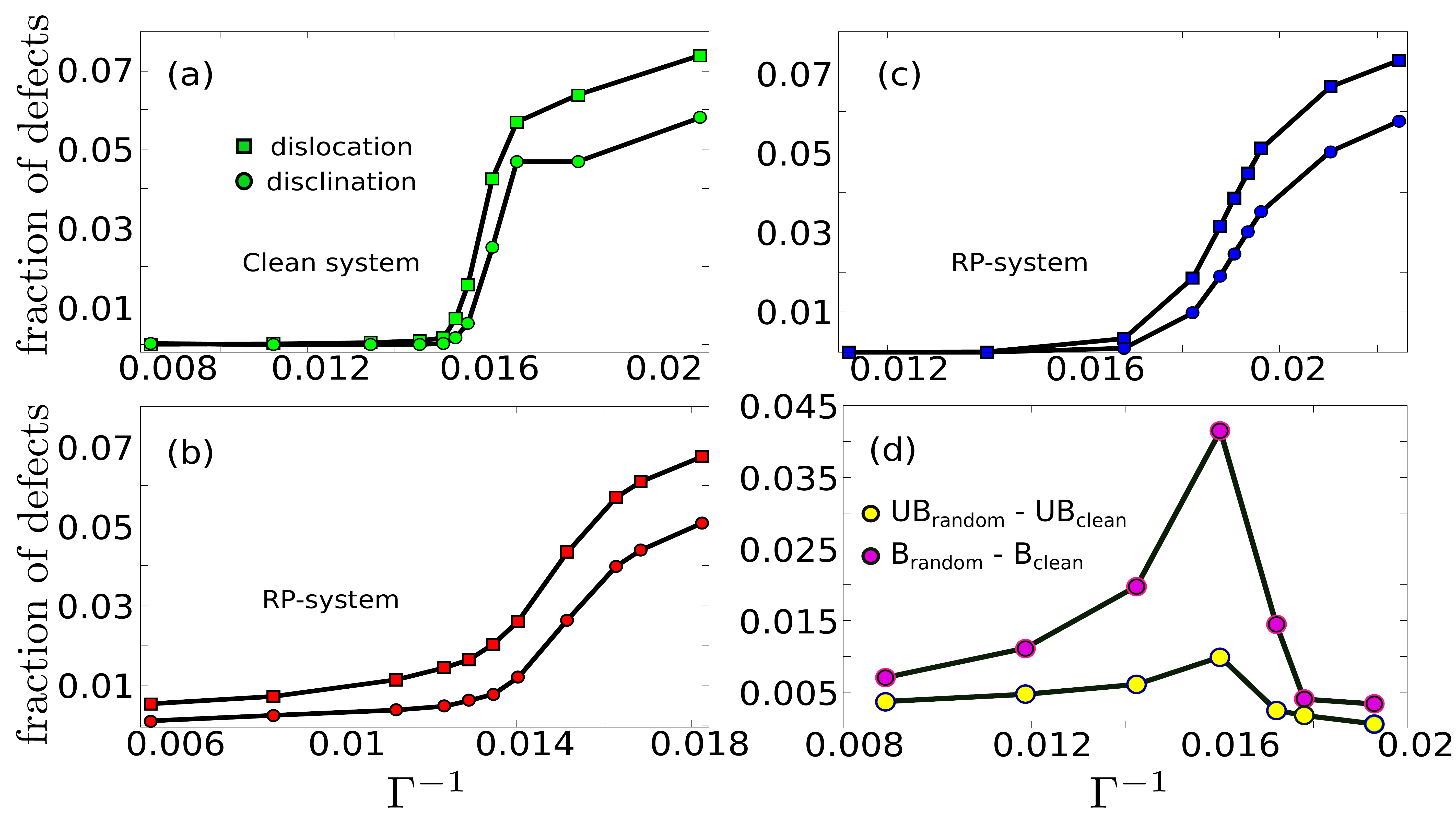}
    \caption{{\bf Evolution of defects:} Panels (a)-(c) represent the variation of defects with $\Gamma^{-1}$ (green: pure, red: RP, and blue: CP-system). (a) The evolution of  unbound dislocations (in squares) as well as disclinations (in circles) validate KTHNY melting in pure systems; (traces of disclinations are multiplied by $10$ for visual clarity). (b) The presence of significant unbound dislocations in RP-systems at low $\Gamma^{-1}$ prohibits solidity. In CP-systems, shown in panel (c), the proliferation of unbound defects (both dislocations and disclinations) commences at a single threshold ($\Gamma^{-1}$=0.0198), implying a direct solid-liquid transition.   Panel (d) shows the comparative number defects in RP-systems with reference to a pure system, illustrating how the number of bound (B) and unbound (UB) defects grow with increasing temperature.}
    \label{f2b}
\end{figure}    
\textit{Defects analysis: \textemdash }
 In KTHNY theory, a 2D-solid transits to the hexatic phase by the unbinding of paired dislocations~\cite{PhysRevLett.41.121, PhysRevB.19.2457}. We proceed to examine the consistency of this picture in Fig.~\ref{f1}. This requires an estimation of the critical distance between two dislocations (with equal and opposite Burger vectors) below which they are bound. We first employ the Hungarian algorithm~\cite{kuhn1955hungarian}, which chooses `correct' partners of dislocations by minimizing the sum of the distances between all partners~\cite{PhysRevB.39.7212}. The distribution of resulting pair distances, $P(r_{\rm dp})$, at low-$T$ is presented in Fig.~\ref{f2}(a-c).
$P(r_{\rm dp})$ is sharply peaked for pure systems in Fig.~\ref{f2}(a), with insignificant weight at larger $r_{\rm dp}$. 
In contrast, its long tail at low $T$ (shown for $\Gamma^{-1}=0.0056$ in Fig.~\ref{f2}(b)) for the RP-system arises from unbound dislocations even for $T\rightarrow 0$, which destabilize a true solid.
$P(r_{\rm dp})$ in CP-systems (Fig.~\ref{f2}(c)) consists of the initial sharp peak, and nearly no weight for larger $r_{\rm dp}$.
A discernible tail in $P(r_{\rm dp})$ for pure systems develops when dislocation pairs start unbinding. An integrated distribution of $P(r_{\rm dp})$ features a threshold behavior at this transition (see SM  ~\cite{f_note}).\\
To obtain the critical $r_{\rm dp}$ for the pure system, we plot in Fig.~\ref{f2}(d) the difference of these distributions, $\Delta P(r_{\rm dp})$, at temperatures just above and below $\Gamma^{-1}_{\rm SH}$, while the corresponding $P(r_{\rm dp})$-s are shown as the inset. The total positive and negative weights of $\Delta P(r_{\rm dp})$ cancel out, and $r^c_{\rm dp} \approx 2.15a_0$ is identified as the last zero-crossing point. This identification is found robust for $T$'s near $\Gamma^{-1}_{\rm SH}$. A study of distances of disclination pairs yielded a similar critical distance between disclination pairs. Once extracted for the pure system, these critical distances were used for analyzing pinned systems.
Subsequently, we explored the thermal evolution of the defects and their unbinding in Fig.~\ref{f2b}(a-c). For the lowest $T$, defects are essentially absent in the pure system. Unbound disclinations proliferate at $\Gamma^{-1}\approx 0.0162$, whereas dislocations unbind at $\Gamma^{-1}\approx 0.0145$, with a hexatic phase at intervening temperatures~\cite{PhysRevLett.106.235701}, consistent with Fig.~\ref{f1}.
The CP-system (Fig.~\ref{f2b}(c)) behaves like a `better' solid at low-$T$ due to the absence of any free defects up to $\Gamma^{-1} = 0.0182$ beyond that unbound dislocations and disclinations start proliferating at the same $\Gamma^{-1}_{\rm CP}$.
There are significant number of impurity-induced unbound dislocations in the RP-system for $T\rightarrow 0$, as also concluded from Fig.~\ref{f1}.
Here, unpaired dislocations are not only present for all $T$, they even outnumber bound dislocations at low-$T$.
Fig.~\ref{f2b}(d) addresses the role of the impurity induced free defects (at $T\rightarrow 0$) in RP-systems, on the thermal defects, whose unbinding drives the two transitions in a pure system. Number of bound (B) and unbound (UB) defects, with corresponding numbers subtracted for an equivalent pure system, are examined separately in Fig.~\ref{f2b}(d).
 These numbers increase sharply with $T$ until the system transits to the liquid. Thus, the impurity induced defects help promote further thermal defects, than in pure systems for  $\Gamma^{-1} < \Gamma^{-1}_{\rm HL}$. Such a rise disappears in the liquid.
 In fact, this difference in bound defects in the liquid goes down to a even lower value than the corresponding number at $T\rightarrow 0$.  \\
 \begin{figure}[t]
    \includegraphics[width=0.48\textwidth]{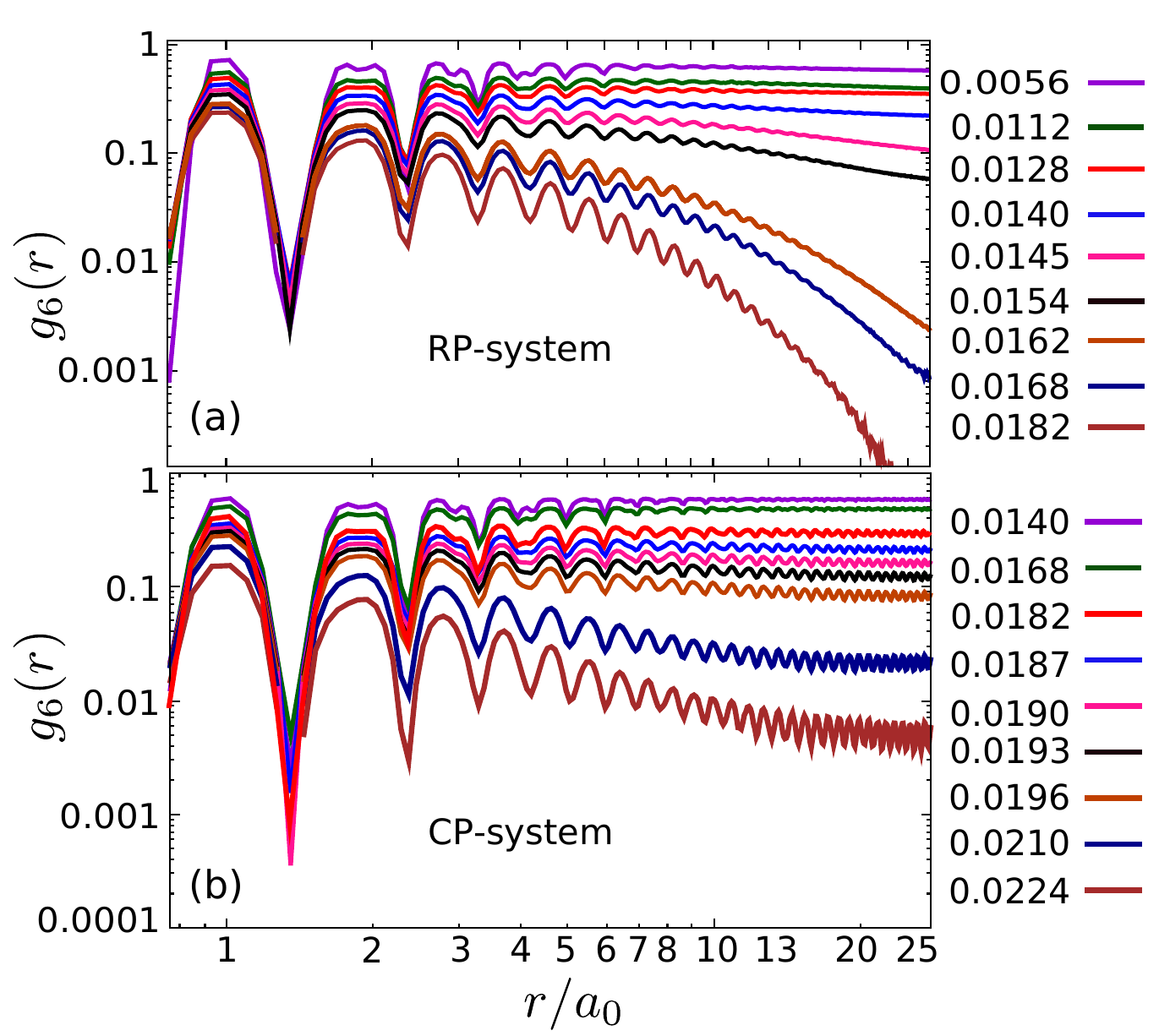}
    \caption{{\bf Correlations:} Panels (a), (b) show the orientational correlation function, $g_{6}(r)$, of  RP- and CP-systems, respectively, for the values of $\Gamma^{-1}$ as labelled in each case.  In panel (a), power law decay of the correlations is visible (for $\Gamma^{-1} < {}^{\rm RP}\Gamma^{-1}_{\rm m}$) down to a very low $T$ in the RP-system; e.g. we find $g_6(r,\Gamma^{-1}=0.0112) \sim (r/a_0)^{-0.14}$, $g_6(r,\Gamma^{-1}=0.0.014) \sim (r/a_0)^{-0.2}$, $g_6(r,\Gamma^{-1}=0.0154) \sim (r/a_0)^{-0.64}$, while $g_6(r,\Gamma^{-1}=0.0162) \sim {\rm exp}^{-0.16(r/a_0)}$, as shown via dashed lines. In contrast, in CP-system, as shown in (b), correlations get modified. Even in the liquid phase, the conventional exponential decay flattens out at large $r$, implying a ``remnant crystallinity" (See text). 
    }
    \label{f3}
    \end{figure}

\textit{Correlations: \textemdash }
Finally, we discuss the orientational correlations as measured by the 
correlation function
$g_{6}(r)=\langle \Psi_{6}({\bf r}_{i})\Psi_{\rm 6}^*({\bf r}_{j}) \rangle$ where $r= \lvert {\bf r}_i -{\bf r}_j\rvert$.
The $T$-dependence of orientational correlations in the pure system follows KTHNY scenario~\cite{PhysRevLett.41.121, PhysRevB.19.2457, PhysRevB.19.1855} as claimed earlier~\cite{PhysRevLett.106.235701}. The evolution of $g_{6}(r)$ for various $T$ is shown in Fig.~\ref{f3} for RP- and CP-systems.
Our $\chi^2$-minimization analysis~\cite{levenberg1944method} of the large-$r$ decay of $g_{6}(r)$ in the RP-system (Fig.~\ref{f3}(a)) identified a power-law behavior for nearly the entire low-$T$ phase.
This power-law behavior continues until an exponential decay sets in for $\Gamma^{-1} \geq 0.0162$, signaling the onset of liquidity.
Intriguingly, $g_{6}(r)$ in CP-systems shows the enhanced solidity for $\Gamma^{-1} \leq 0.0187$, where its traces remain largely flat.
Beyond the direct melting from solid to liquid for $\Gamma^{-1} > 0.0190$, $g_{6}(r)$ in CP-systems displays a tendency of plateauing at large-$r$, though it decays at intermediate $r$. This is a signature of `remnant solidity' arising from local crystalline pockets surrounding the impurities
 whose locations are  commensurate with the perfect crystal and hence anchoring crystallinity in the vicinity (See SM  ~\cite{f_note}).
This is a direct consequence of the correlated nature of the CP-impurities. \\
\textit{Conclusion: \textemdash }
To summarize, we demonstrate that the conventional picture of 2D-melting undergoes significant changes in the presence of impurities.
While RP-disorder destabilizes solidity and CP-disorder removes the hexatic phase, the low-$T$ phase in RP-systems is not the conventional hexatic. Similarly, the high-$T$ phase in the CP-systems mixes remnant solidity with the liquid phase.
The inhomogeneous melting (Fig.~S1 in SM ~\cite{f_note} generates defects which correlate differently with pinning centers:
For RP-systems the defects tend to bind with the pinning centers, whereas the defects stay away from the impurities in CP-systems. 
 While defects are found essential for driving the melting, our MD configurations indicate that they often bunch up in various shapes of macroscopic size (See videos in SM~\cite{f_note}).
An extension of our study to larger systems exploring possible role of grain boundaries on melting is a promising future direction. It will also be interesting to inspect the role of quantum fluctuations in these thermal phases. We hope that our findings will motivate future experiments for shedding new light.

\nocite{}

\bibliography{ref}

\begin{thebibliography}{51}
\bibitem [{f_n()}]{f_note}%
  \BibitemOpen
  \href@noop {} {}\bibinfo {note} {The URL for SM will appear here}\BibitemShut
  {NoStop}%
\end{thebibliography}
\clearpage  
\onecolumngrid  
\begin{center}
\textbf{\LARGE Supplementary material for `The effect of disorder on phases across two-dimensional thermal melting'}
\end{center}

In order to support the key conclusions reported in the main manuscript, we include below additional results in this supplementary materials (SM).

\section{Model and Methods}

For our study, we consider the Gaussian-core model in a 2D system with periodic boundary conditions, within which particles act via the following interaction. $V(r) = \epsilon exp(-r^2/\sigma^2)$, with $\epsilon>0$, $\epsilon$ and $\sigma$ are the energy and length scales, respectively.
We carry out  molecular dynamics (MD) simulations of this model system, using LAMMPS, starting with a high temperature and cooling the system down to the desired temperature, as detailed in the main text.
We use $10^5$ MD steps ($t = 5000$) with sampling time window of $t=0.1$, which are recorded after $10^7$ sweeps of equilibration runs. The desired temperatures are maintained via the Berendsen thermostat. 
To ensure correct equilibration before taking statistics, we
looked into the distribution of velocities of the particles, which assumes the form of an Maxwell-Boltzmann
distribution in thermal equilibrium as well as independence of the temporal correlation of the observables on the time origin, i.e. $\zeta(t_{1},t_{2}) = \zeta_(t_{1}-t_{2})$, where $\zeta$ is any temporal correlation defined between two time points $t_{1}$ and $t_{2}$.
\setcounter{figure}{0}  
\renewcommand{\thefigure}{S\arabic{figure}}  
\begin{figure}[t]
\includegraphics[width=1\textwidth]{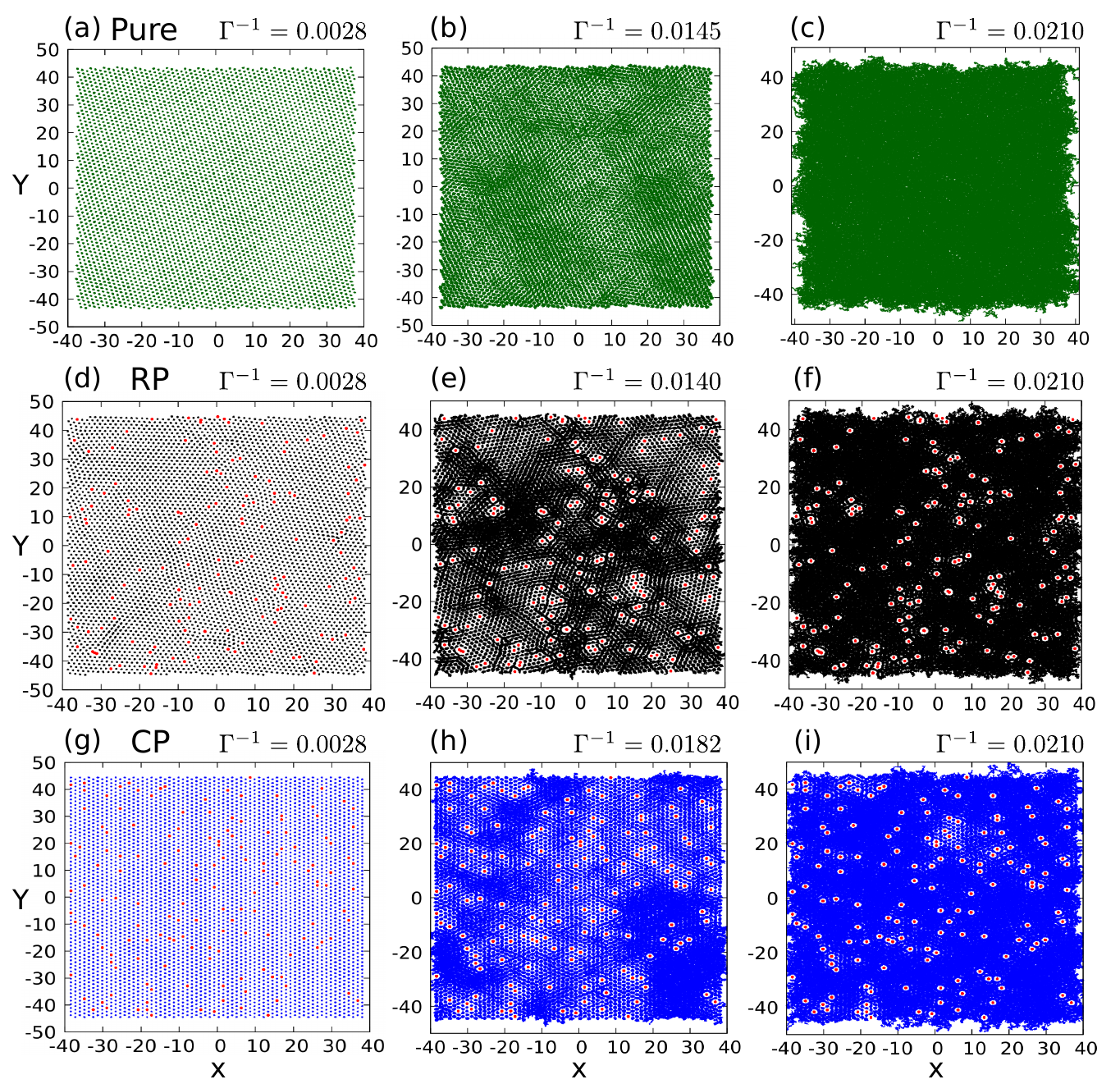}
\caption{\textbf{Trajectories:} The trajectories
 of the particles studied at different temperatures ($\Gamma^{-1}$) for all three systems (the pinning centers marked in red) for $\Delta t = 5000$. Panels (a-c) represent the trajctories of particles in the clean system. For lowest $\Gamma^{-1}$ (panel a), the particles remain firmly tied to the equilibrium position of a triangular lattice reflecting the solid phase. In the hexatic phase at $\Gamma^{-1}=0.0145$ (panel b), as per Fig.~2 of the main text, particles tend to move preferentially along the principal direction.
 In the liquid phase (panel c), particles execute isotropic motion, covering the entire space.
 The RP-system at low-$T$ (panel d) features distorted lattice lines and a resulting irregular array of particles. Together with the results from Fig.~2 of main text, the trajectory of particles indicate that the lowest-$T$ phase of RP-system constitute a hexatic glass.
 Panel (e) depicts the trajectory plot near ${}^{\rm RP}\Gamma^{-1}_{\rm m}$ and features incipient melting which results into motional signatures which are rather inhomogeneous in space. Finally, the liquid phase in the RP-system in panel (f) displays particles' motion everywhere except for a small region surrounding the impurities.
The motion in CP-system looks similar to the clean system in the solid phase (panel g). Close to the ${}^{\rm CP}\Gamma^{-1}_{\rm m}$ in panel (h) the dynamics shows inhomogeneity, but unlike in RP-system the mobile particles tend to stay away from pinning centers.
the concentration of pinned centers keeps away the melting from the pinned centers (panel h).
While most particles delocalize significantly in the high-$T$ phase of the CP-system (panel i), local patches of solid-like regions still survive, as seen from the trajectory plot.}
\label{gcm_s1}
\end{figure}

\section{Snapshots of particle trajectories}
Here we plot the trajectories of particles during their equilibrium dynamics for the clean, RP- and CP-systems. Such motional signatures are displayed in Fig.\ref{gcm_s1} for three representative temperatures, one at low-$T$ in the solid phase (likely a hexatic glass in case of RP-system), at intermediate temperature (just before the system melts into an isotropic liquid (in hexatic phase in case of the clean system), and finally at a high temperature (deep into the liquid state). 
The following points are worth mentioning:\\
\noindent The results for clean system in Fig.~\ref{gcm_s1}(a-c) are consistent with the KTHNY picture of 2D-melting in a pure systems. Note that the particles in a hexatic phase (Fig.~\ref{gcm_s1}(b)) preferentially move along the three principal directions of the underlying triangular lattice of the solid.\\
\noindent The lowest temperature phase for the RP-system in Fig.~\ref{gcm_s1}(d) represents a hexatic glass (or, an amorphous solid) in which the pinned impurities are marked red, with no apparent positional order. The snapshot at $\Gamma^{-1}=0.0140$ for RP-system in Fig.\ref{gcm_s1}(e) appears inhomogeneous, where the mobile particles close to pinned particles delocalize more. Other particles tend to carry the motional signature of a hexatic phase in a clean system. The snapshot in Fig.\ref{gcm_s1}(f) for RP-system for $\Gamma^{-1}=0.0210$ indicates that delocalization of mobile particles is nearly engulfed the whole space, except for small region  surrounding the repulsive pinning centers.\\
\noindent The snapshots in the CP-system in Fig.\ref{gcm_s1}(g-i) supports the notion of enhanced solidity in the following manner: The solidity is nearly perfect at low $T$ (solidity here is stronger than in clean system) -- the impurities are pinned at commensurate locations of the underlying triangular lattice hold onto perfect solidity. Unlike for RP-system, here the regions rich in pinning centers anchors solidity around them, whereas, the regions relatively free of impurity feature incipient melting. This is seen from Fig.\ref{gcm_s1}(h) for $\Gamma^{-1} = 0.0182$, i.e. at a temperature just below the onset of melting. As a result, the CP-system displays a local pockets of `remnant solidity' around impurities. This remnant solidity weakens with $\Gamma^{-1}$, however, persists up to a large temperature (Fig.\ref{gcm_s1}(i) at $\Gamma^{-1} = 0.0210$). Note that the pinning centers maintain solid-like correlation at all $T$.
The remnant solidity has discernible effects on the bond-orientational correlations $g_{\rm 6}(r)$, discussed in connection with Fig.~5(d) in the main manuscript, and is elaborated further in the later part of this SM.

\section{Identification of critical distances of defects unbinding}

\begin{figure}[t]
\includegraphics[width=1\textwidth]{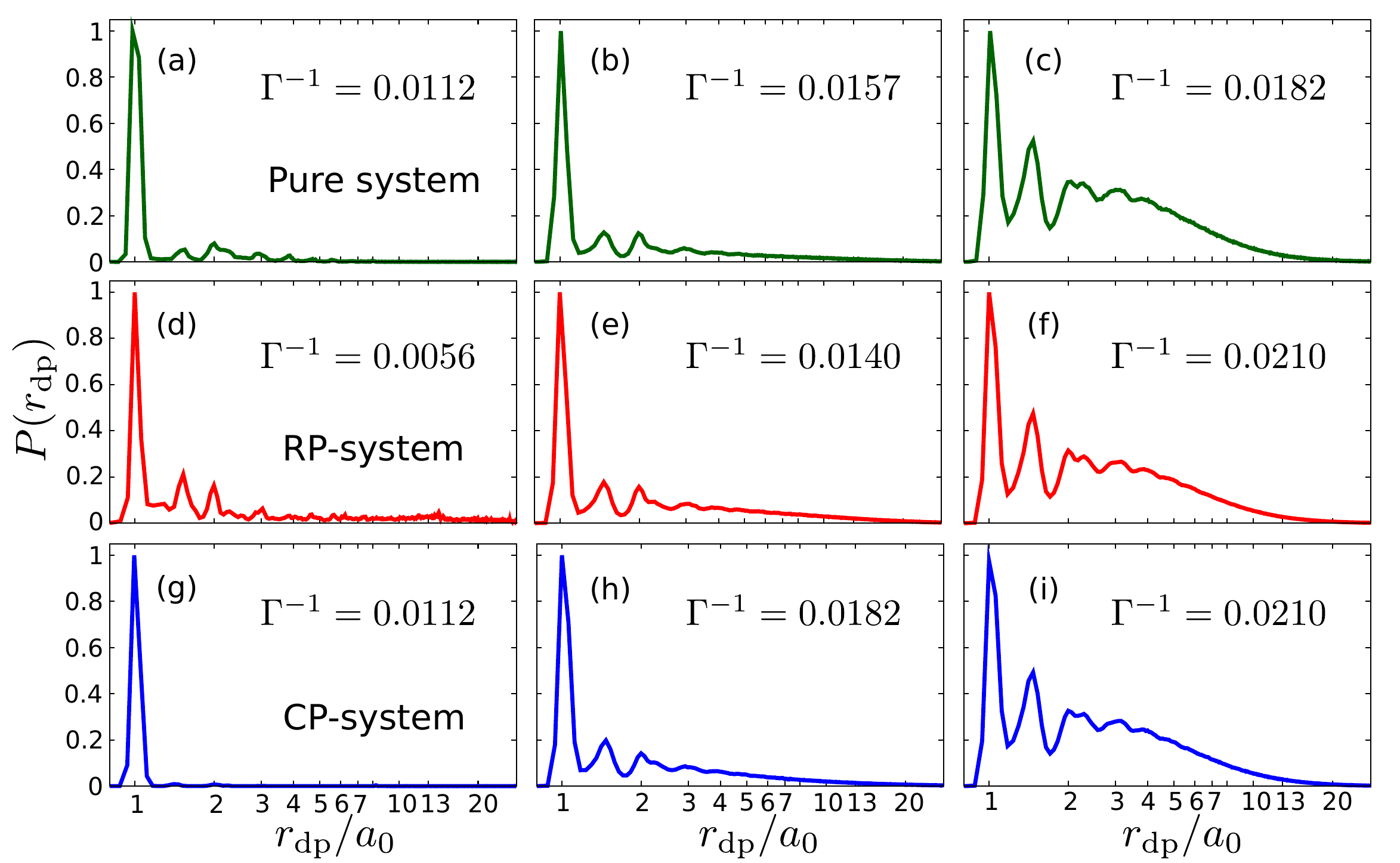}
\caption{\textbf{ \bf The distribution of distances between dislocation pairs:} $P(r_{\rm dp})$ shown for different phases. Panels
(a-c), (d-f), and (g-i) show results for clean, RP-, and CP-systems, respectively. $P(r_{\rm dp})$  at low-$T$ in panels (a), (d), and
(g) are sharply peaked at the typical distance of dislocation pairs, though there are differences in details. 
The distribution has a short tail for the clean system, and a very long
tail for RP-systems, and essentially no tail for CP-systems, indicating
the enhanced solidity due to commensurate pinning.
The distributions for the three systems develop smooth and healthy tails
at a $T$, where defects unbinding sets in, as shown in panels (b, e, h). Panels (c, f, i) show that they become indistinguishable at large $T$, beyond the melting to a liquid.}
\label{defct_distn}
\end{figure}
In fig.~\ref{defct_distn}, we report the distribution of the separation of the two dislocations with equal and opposite Burger's vector (here the two constitute a bound pair of dislocations). We denote this distribution by $P(r_{\rm dp})$, where $r_{\rm dp}$ is the separation in question.
At lower $T$ ($\Gamma^{-1}=0.0112$), Fig.~\ref{defct_distn}(a) show that $P(r_{\rm dp})$ in a clean system is sharply peaked at 
$r_{\rm dp} \approx a_{0}$, the average inter-particle distance, with some additional weight for larger $r_{\rm dp}$. In contrast, a weak yet very long tail in $P(r_{\rm dp})$ (Fig.~\ref{defct_distn}(b)) is found for the
RP-system at low-$T$. This demonstrates the presence of unbound dislocations even at the lowest $T$ ($\Gamma^{-1}=0.0056$), prohibiting true solidity.  In contrast, the first peak in $P(r_{\rm dp})$ for CP-systems at low-$T$ in Fig.~\ref{defct_distn}(g) is even sharper than the one in clean system (Fig.~\ref{defct_distn}(a)). This reflects an enhanced solidity in the CP-system. The nature of the $P(r_{\rm dp})$ changes with $T$, and
when the dislocation pairs start unbinding, $P(r_{\rm dp})$ begins to develops smooth tails, see Fig.~\ref{defct_distn}(b, e, h). At these $T$'s and beyond $P(r_{\rm dp})$ becomes qualitatively similar for clean, RP- and CP-systems. The distributions become nearly identical deep into the liquid phase (hight-$T$) as shown in Fig.~ \ref{defct_distn}(c, f, i).

\begin{figure}[t]
\centering
\includegraphics[width=0.95\textwidth]{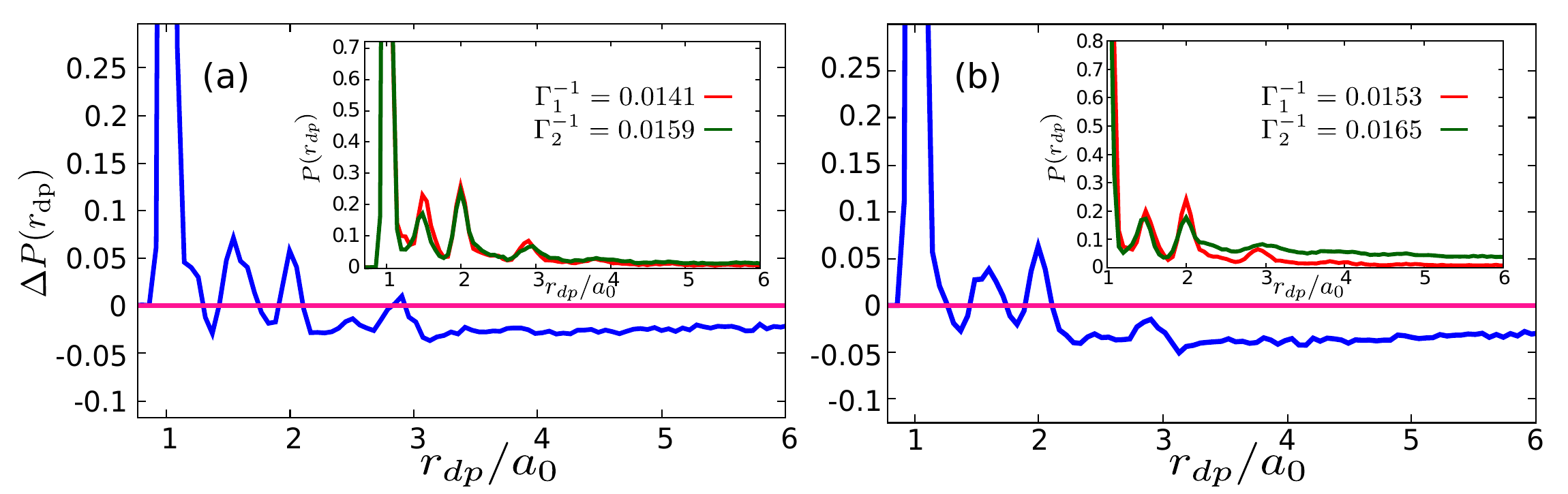}
\caption{\textbf{Difference in \boldmath {$P(r_{\rm dp})$:}} The two panels display $\Delta P(r_{\rm dp})$ in the pure system, taken between
two pair of temperatures -- one in the solid and the other in the hexatic phase, both being close to ${}^{C}\Gamma^{-1}_{\rm SH}$. The corresponding traces of $P(r_{\rm dp})$ is shown as the
inset. The largest $r_{\rm dp}$ corresponding to a zero-crossing is taken as $r_{dp}^c$.}
\label{diff_pdr}
\end{figure}

Fig.~\ref{diff_pdr} supplements the Fig.~3(d) in the main text, and illustrates the insensitivity of $r_{dp}^c$ on the chosen $T$-values for its extraction, as long as those temperatures are close to ${}^{C}\Gamma^{-1}_{\rm SH}$. We plot in Fig.~\ref{diff_pdr} the difference $\Delta P(r_{\rm dp})$ of $P(r_{\rm dp})$ for two different pair of temperatures around the ${}^{C}\Gamma^{-1}_{\rm SH}$, while the corresponding distributions are shown as the inset. Both plots yield  $r_{dp}^c = 2.15a_{0}$ -- the same value obtained in the main test for a yet another pair of $T$'s. 
\begin{figure}[t]
\centering
\includegraphics[width=0.6\textwidth]{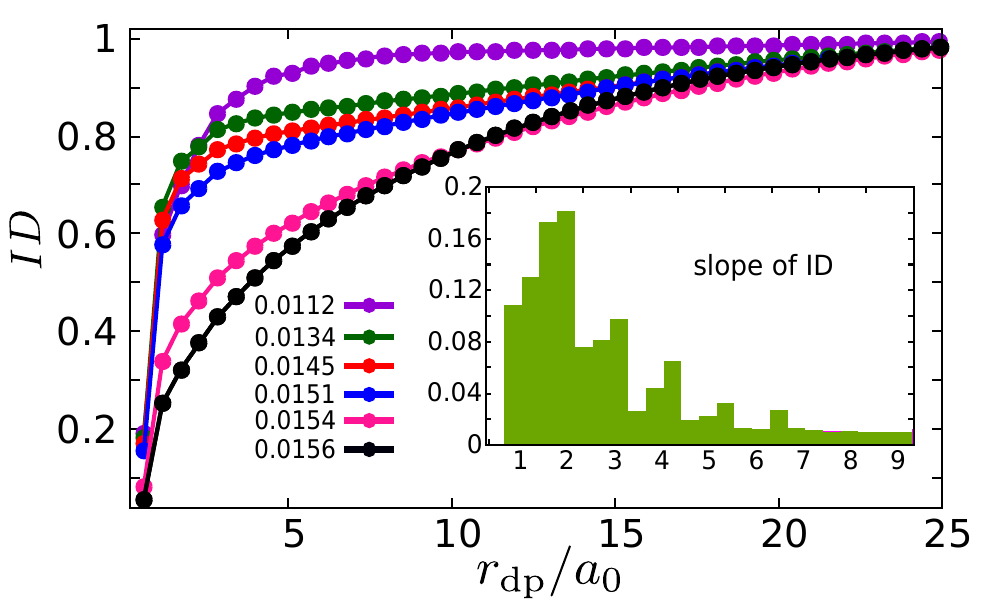}
\caption{\textbf{integrated distribution (ID):} as a function of $r_{\rm dp}$ for different $Ts$. The attainment of unity of ID occurs sharply in the solid phase and more gently beyond ${}^{C}\Gamma^{-1}_{\rm SH}$. The curvature of ID is studied further to estimate the critical distance ($r_{dp}^{c}$), which is presented as an  inset.}
\label{ipd}
\end{figure}
An alternative extraction of $r_{\rm dp}^{c}$ for the pure system, employing differently the same concept as above is the following:
In Fig.~\ref{ipd}, we plot the integrated distribution (ID) of $P(r_{\rm dp})$ for various $T$'s. The traces of ID shows a threshold behaviour marking the solid-to-hexatic transition. ID comes up very steeply for the lower $T$'s corresponding the solid phase (as inerred from the Fig.~2 of the main text), whereas the rise of ID becomes distinctively gradual for the two higher $T$'s representing a hexatic phase. In the inset, we plot the distribution of the change of the slope of ID-traces in the solid phase, which attains the peak-value for $r_{\rm dp}^{c} \approx 2.15a_{0}$.

\begin{figure}[t]
\includegraphics[width=1\textwidth]{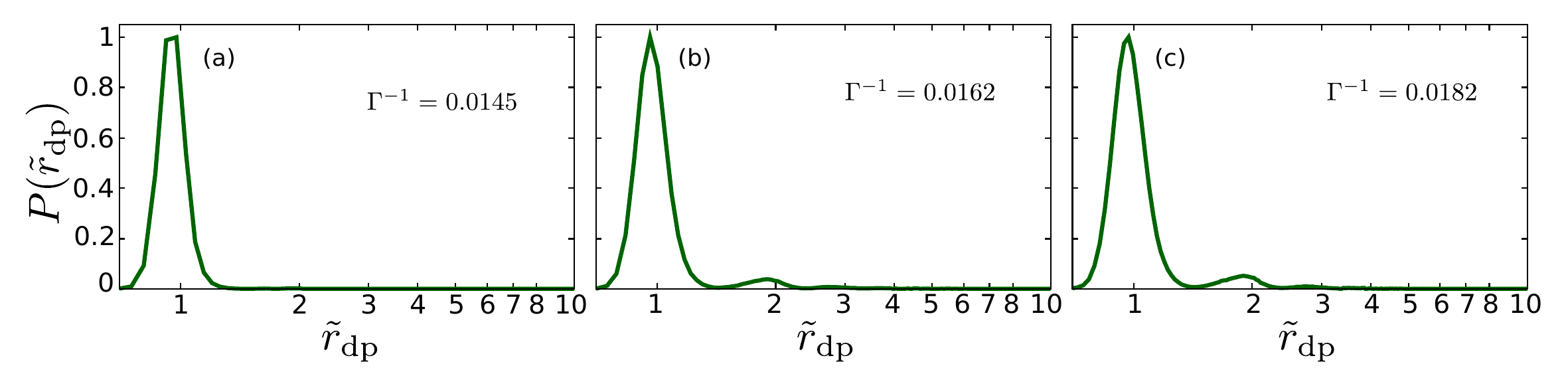}
\caption{\textbf{ distribution of distances between disclination pairs:} Panels (a-c) represent the distribution of the distances of the disclination pairs $P({\tilde r}_{\rm dp})$ at different $\Gamma^{-1}$. Panel (a) is the $P({\tilde r}_{\rm dp})$ corresponds to the hexatic phase, which attains a sharp peak around the lattice spacing reflecting the only presence of tightly bounded disclinations in the phase. When the system hits the transition $T$, the defects unbounds. The unbinding of these defects modifies the distribution, as shown in panels (b) in the liquid phase, and remains identical for the higher $\Gamma^{-1}$ value (panel (c)).   }
\label{defct_discln}
\end{figure}

We also studied the thermal evolution of the distribution of the disclination pair (${\tilde r}_{\rm dp}$) and we present our results in fig~.\ref{defct_discln} for three representative $T$'s. This distribution $P({\tilde r}_{\rm dp})$ features a single peak at ${\tilde r}_{\rm dp}\sim a_0$ for up to ${}^C\Gamma^{-1}_{\rm HL}$, and identify a second peak for larger $T$'s at ${\tilde r}_{\rm dp}\approx 2a_0$. Thus, we conclude that ${\tilde r}^c_{\rm dp} \sim r^c_{\rm dp}$.

\section{Correlations:}
\begin{figure}[t]
\includegraphics[width=1\textwidth]{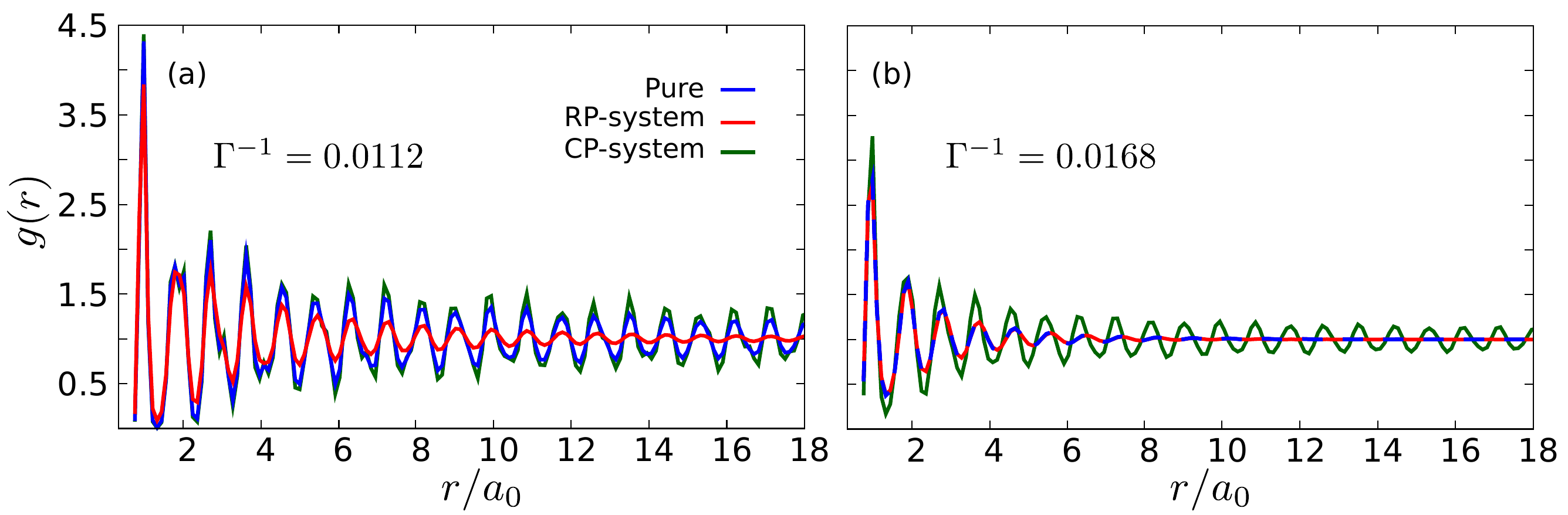}
\caption{\textbf{Pair correlation function:} Panel (a) represents the pair correlation (red: clean-, blue: random- and  green: CP-system) for two different $\Gamma^{-1}$ values. At the low $\Gamma^{-1}$ value, the clean system's behavior is comparable to the CP-system, while in the higher $\Gamma^{-1}$, as shown in panel-b (for visual clarity, the traces of RP-system is shown in a dotted line), it is comparable to the RP-system. }
\label{gofr}
\end{figure}

The pair correlation is given by 
\begin{equation}
    g(r)=\frac{1}{2\pi rN} \sum_{i=1}^{N}\sum_{j\neq i=1}^{N}
\langle \delta(r-\lvert\bf{r}_{i}-\bf{r}_{j}\rvert) \rangle , 
\end{equation}
 with $r= \lvert {\bf r}_i -{\bf r}_j\rvert$. We present the low-$T$ behavior of $g(r)$ in Fig.~\ref{gofr}(a). Its evolution in the clean (blue) and CP (green) systems is nearly identical, consistent with the findings of Fig.~2(a) and Fig.~2(b) in the main text. The $g(r)$ for the RP-system exhibit damped modulations which prohibits solidity even at low temperatures. However, these damped modulations are long-ranged than what's expected in a liquid.
Fig.~\ref{gofr}(b) shows $g(r)$ for large $T$ where all three systems turns into `liquid.' A role reversal occurs -- $g(r)$ for clean and RP-system overlap this time, yielding a liquid-like behavior, while the long-range modulations for CP-systems survives implying `remnant solidity' as discussed already in other context. 

\begin{figure}[t]
\centering
\includegraphics[width=0.75\textwidth]{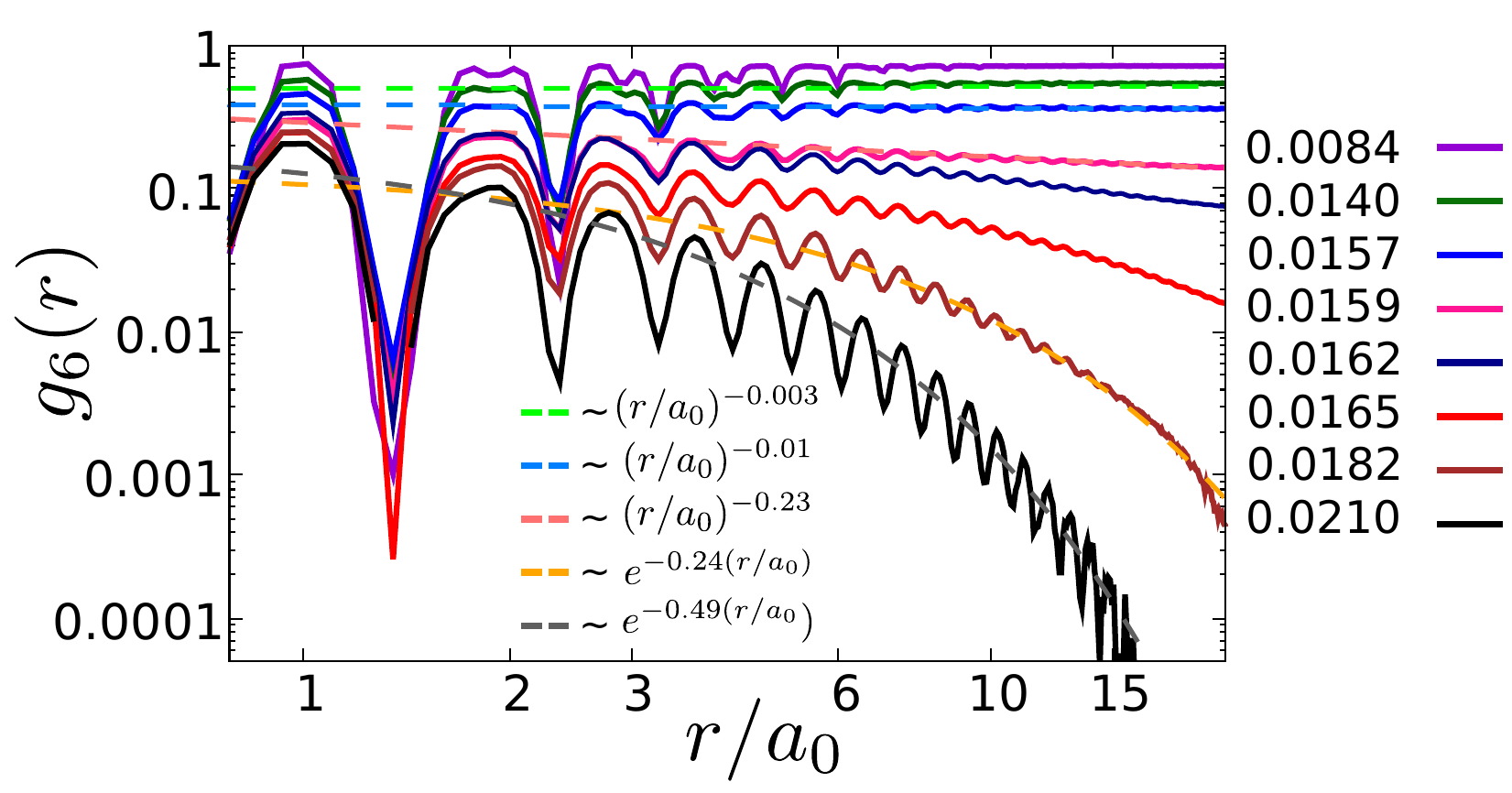}
\caption{\textbf{ Orientational correlation function:} of a clean system at different $Ts$. The pure system displays the thermal evolution of $g_6(r)$ consistent with KTHNY theory.   }
\label{cln_g6}
\end{figure}
The $T$-dependence of orientational correlations in the clean system is shown in Fig.~\ref{cln_g6} implies that $g_{6}(r)$ remains nearly independent of $\Gamma^{-1}$ for up to $\Gamma^{-1}= 0.0145$. It shows a power-law decay $\sim r^{-\eta(T)}$, in the hexatic phase for $0.0151 \geq \Gamma^{-1} \geq 0.0162$ (exponents $\eta(T)$ are listed in Fig.~\ref{cln_g6}). Finally, $g_{6}(r)$ features an exponential decay in the liquid phase for $\Gamma^{-1} \geq 0.0165$. We find, e.g., $g_6(r,\Gamma^{-1}=0.0140) \sim (r/a_0)^{-0.003}$, $g_6(r,\Gamma^{-1}=0.0.0157) \sim (r/a_0)^{-0.01}$, $g_6(r,\Gamma^{-1}=0.0159) \sim (r/a_0)^{-0.23}$, while $g_6(r,\Gamma^{-1}=0.0182) \sim {\rm exp}^{-0.24(r/a_0)}$ and $g_6(r,\Gamma^{-1}=0.0210) \sim {\rm exp}^{-0.49(r/a_0)}$.

\begin{figure}[t]
\centering
\includegraphics[width=0.6\textwidth]{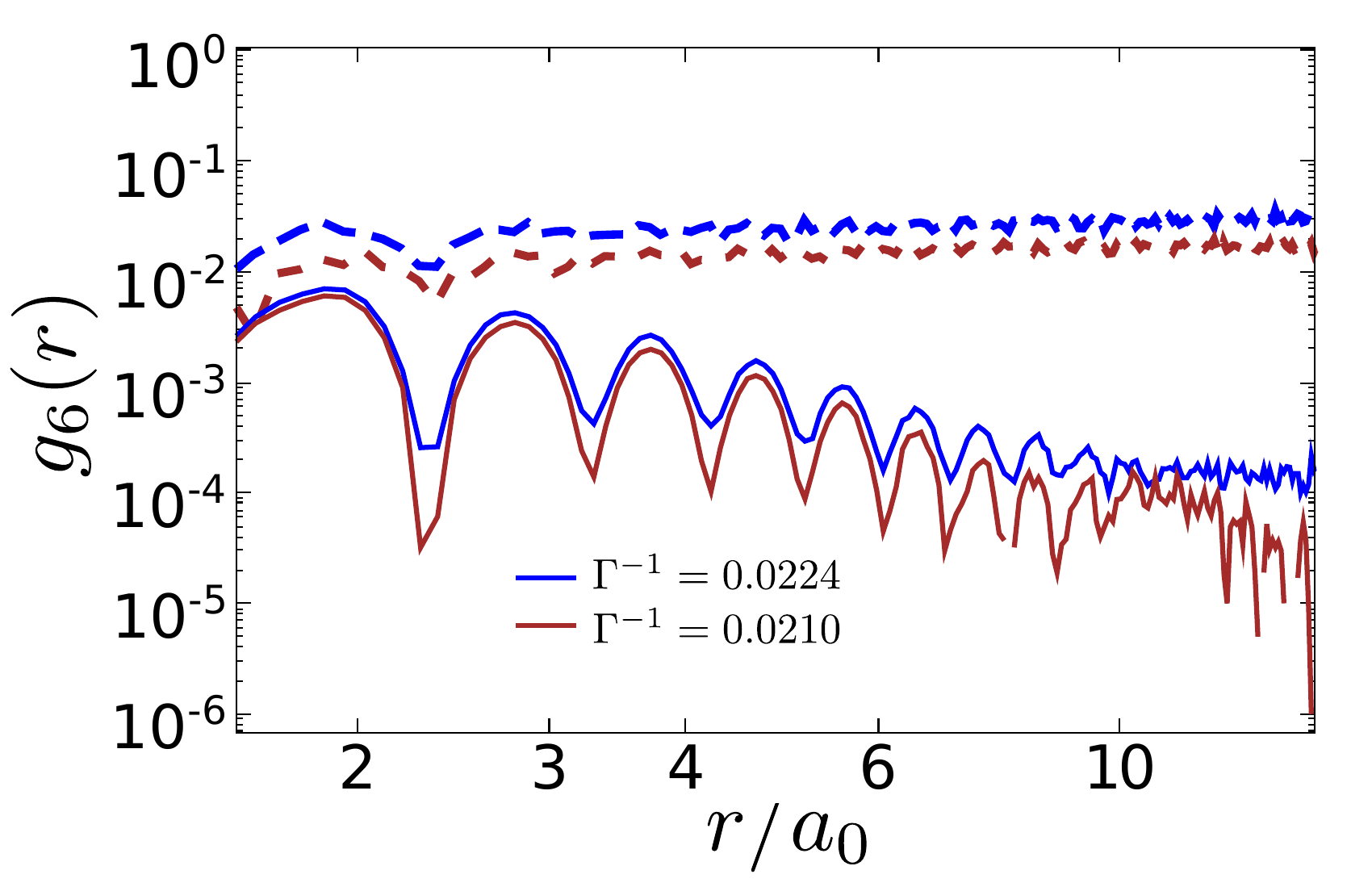}
\caption{\textbf{ Orientational correlation function,} The $g_{\rm 6}(r)$ is calculated in two separate components. (i) The dotted line shows the contribution of $g_{\rm 6}(r)$ only for the particles which lie within a distance of $2a_0$ from pinned particles at the higher $Ts$ beyond the melting temperature and (ii) the solid traces shows the  $g_{\rm 6}(r)$ contributed by all other particles which are more than $2a_0$ distance.} 
\label{cp_g6}
\end{figure}

In order to address `remnant solidity' in the CP-system at high $T$, we plot the two separate components of $g_{\rm 6}(r)$ in Fig.~\ref{cp_g6}. Once component estimates the contribution of $g_{\rm 6}(r)$ (dotted trace) only for the particles which lie within a distance of $2a_0$ from pinned particles. The other component measures $g_{\rm 6}(r)$ (solid trace) contributed by all other particles. We present data of such two contributions for two values of $T$ (both in the high-$T$ liquid phase). The message from Fig.~\ref{cp_g6} is clear -- the particles close to the pinning centers at commensurate locations of a perfect crystal anchor the crystallinity around them, which is reflected in the corresponding $g_{\rm 6}(r)$ which remains nearly constant at large $r$ like in a solid. The particles further away from the pinned particles, on the other hand, expectedly behaves like a liquid. Thus, it is the particles near pinning centers which contribute to the `remnant solidity.' As such, that pinned particles must show a perfect solid like correlation by construction!

\end{document}